\documentclass[aps,pre,reprint, amsmath, amssymb,superscriptaddress,nofootinbib]{revtex4-1}
\usepackage[utf8]{inputenc}

\usepackage{natbib}
\usepackage{graphicx,xcolor}
\usepackage{amsmath,mathtools}
\usepackage{amssymb}
\usepackage{physics}
\usepackage{enumerate}
\usepackage{float}
\usepackage{mathtools}
\usepackage{bbold}
\usepackage{comment}
\usepackage{subcaption}
\usepackage{dsfont}
\usepackage{ulem}

\usepackage{hyperref}
\definecolor{darkred}  {rgb}{0.5,0,0}
\definecolor{darkblue} {rgb}{0,0,0.5}
\definecolor{darkgreen}{rgb}{0,0.5,0}
\hypersetup{
	colorlinks = true,
	urlcolor  = blue,         
	linkcolor = darkblue,     
	citecolor = darkgreen,    
	filecolor = darkred       
}

\newcommand{\ba}{\begin{eqnarray}}
\newcommand{\ea}{\end{eqnarray}}
\newcommand{\ban}{\begin{eqnarray*}}
\newcommand{\ean}{\end{eqnarray*}}

\newcommand{\set}[1]{\mathcal{#1}}
\renewcommand{\>}{\rangle}
\newcommand{\<}{\langle}
\newcommand{\mE}{\mathcal{E}}

\newcommand{\aver}[1]{\left\langle #1 \right\rangle}

\begin{document}
\title{Fluctuation theorems from Bayesian retrodiction}

\author{Francesco Buscemi}
\email{buscemi@nagoya-u.jp}
\affiliation{Graduate School of Informatics, Nagoya University, Chikusa-ku, 464-8601 Nagoya, Japan}

\author{Valerio Scarani}
\affiliation{Centre for Quantum Technologies, National University of Singapore, 3 Science Drive 2, Singapore 117543, Singapore}
\affiliation{Department of Physics, National University of Singapore, 2 Science Drive 3, Singapore 117542, Singapore}

\begin{abstract}
Quantitative studies of irreversibility in statistical mechanics often involve the consideration of a reverse process, whose definition has been the object of many discussions, in particular for quantum mechanical systems. Here we show that the reverse channel very naturally arises from Bayesian retrodiction, both in classical and quantum theories. Previous paradigmatic results, such as Jarzynski's equality, Crooks' fluctuation theorem, and Tasaki's two-measurement fluctuation theorem for closed driven quantum systems, are all shown to be consistent with retrodictive arguments. Also, various corrections that were introduced to deal with nonequilibrium steady states or open quantum systems are justified on general grounds as remnants of Bayesian retrodiction. More generally, with the reverse process constructed on consistent logical inference, fluctuation relations acquire a much broader form and scope.
\end{abstract}

\date{\today}

\maketitle


\section{Introduction}

In modern statistical mechanics, it has become customary to capture irreversibility by a suitable comparison between a forward and a backward (or reverse) process. Such a comparison is formulated in terms of \textit{fluctuation relations}. Initially limited to linear response, such relations have progressively been extended to encompass a much larger class of processes~\cite{bochkov-kuzovlev-1977}. After the widely noticed works of Jarzynski~\cite{jarzynski} and Crooks~\cite{crooks-theorem}, the literature has grown at such a fast pace that we can only point the reader to some reviews on the matter~\cite{jarzynski-review-2011,campisi-haenggi-review-2011,gawedzki2013fluctuation,Funo2018}.

Forward and backward processes include a prior (initial state), which can be chosen arbitrarily, and a transition rule, viz. a channel. The forward process is the ``physical process,'' i.e., the propagation of the prior through the physical channel: as such, its definition is unproblematic. The main focus of this paper is the construction of the reverse process. For some processes, the identification is clear: for instance, in the case of classical Hamiltonian dynamics, the reverse transition can be easily identified with the dynamics that generates the time-reversed trajectory. For some other processes, however, physical intuition may not be sufficient. Notably, when the transition rule is governed by an underlying quantum channel, there is currently a widespread belief (see e.g.~\cite{Albash2013,rastegin-zycz-2014,Funo2018}) that Crooks' fluctuation theorem and Jarzynski's equality hold without modifications or other corrections only when the evolution is unitary (i.e., closed) or at least unital (i.e., preserving the uniform distribution). However, non-unital channels such as partial swaps~\cite{ziman02,Lorenzo,strasberg17} or thermal processes~\cite{horo-oppen-thermal} are much more obvious paradigms of thermodynamic irreversibility than unital channels. Besides, classical fluctuation theorems are known to hold also for processes that are surely not unital: now, quantum theory should be a generalisation of classical probability theory, not a restriction over it. 

In this paper, we propose that the reverse transition can be systematically constructed as a form of \textit{Bayesian retrodiction}~\cite{watanabe55,jeffrey,pearls,jaynes_2003,CHAN200567,jacobs-changing-mind}. This retrodictive structure can be recognized in some of the most famous classical fluctuations relations~\cite{bochkov-kuzovlev-1977,jarzynski,crooks-theorem,jarz2000,hatano-sasa-2001}, even if their original derivation was based on different arguments. Then we prove that, when the forward process is realised as a quantum process (preparation, evolution, and measurement), the Bayesian-reversed process is always realised as a valid quantum process as well. This process coincides with the ``quantum retrodiction'' independently defined in several works~\cite{barnett-pegg-jeffers,fuchs2002quantum,Leifer-Spekkens,bergou-symm-retro}, and justifies on general grounds the use of Petz's reverse map~\cite{petz} in the context of fluctuation relations. Our construction applies to any choice of preparation states, quantum channels, and final measurements, leading to exact fluctuation relations that do not require any  modification.

The paper is structured as follows. In Section \ref{sec:FT}, as a preparation, we show how fluctuation relations constitute information-theoretic divergence measures between the forward and the reverse process, and we derive a whole new family of fluctuation relations that originates from Csisz\'ar's $f$-divergences~\cite{Csiszar-1967-f-div}. In Section~\ref{sec:Bayes} we introduce the theory of Bayesian retrodiction and derive the general expression for the classical Bayesian-reversed process. In Section~\ref{sec:qinside}, we consider processes with a quantum realization and show that the Bayesian-reversed process precisely coincides with the quantum retrodiction of the quantum forward process. In Section~\ref{sec:examples}, we recast several known classical fluctuation relations in terms of Bayesian retrodiction. Finally, in Section \ref{sec:exquantum}, we consider the most general case of processes arising from arbitrary quantum channels. In this case we show how our approach very naturally accounts for, and thus justifies as an implicit Bayesian inversion, all the corrections that were before introduced to deal with situations such as nonequilibrium steady states and nonunital quantum channels.


\section{Generalized fluctuation relations from information divergences}
\label{sec:FT}

\subsection{Irreversibility as divergence between the forward and the reverse processes}

As mentioned, studies of irreversibility involve a comparison between a forward ($F$) and a reverse ($R$) process~\cite{Seifert-2005}. Consider the evolution of a system from an initial time $t=0$ to a final time $t=\tau>0$. Let us assume, for simplicity, that the system possesses a finite state space $\set{A}$. Suppose now that both forward and reverse processes, respectively, are given to us in terms of two suitable joint probability distributions, $P_F(x,y)$ and $P_R(x,y)$, respectively, where $x\in\set{A}$ labels the state of the system at time $t=0$, while $y\in\set{A}$ labels the state of the system at time $t=\tau$. Concretely, $P_F(x,y)$ denotes the probability that the system starts in state $x$ at time $t=0$ and ends in state $y$ at time $t=\tau$ under the forward process, while $P_R(x,y)$ denotes the probability that the system starts in state $y$ at time $t=\tau$ and ends in state $x$ at time $t=0$ under the reverse process. Notice that at this point we are not yet preoccupied with the problem of determining what makes the two processes one the ``reverse'' of the other. This will be the main issue in the rest of the paper, but for the time being we are just considering $P_F(x,y)$ and $P_R(x,y)$ as given.


If really $P_F(x,y)$ and $P_R(x,y)$ are one the \textit{reverse} of the other, then \textit{irreversibility} may be quantified in terms of how much the two distributions differ, in agreement with the idea that a reversible situation should correspond to the case in which $P_F(x,y)=P_R(x,y)$ for all $x,y\in\set{A}$. In mathematical statistics, the degree of ``disagreement'' of two distributions is captured in terms of information divergences. Here we focus on the family of $f$-divergences, defined as\footnote{In fact, $f$-divergences are usually defined in terms of $1/r$, see e.g.~\cite{liese-miescke}. However, for the sake of the present discussion, formulas are more easily recognizable with the alternative definition.}~\cite{Csiszar-1967-f-div}
\begin{align}
D_f(P_F\| P_R)&:=\sum_{x,y}P_F(x,y)\ f\!\left(\frac{P_F(x,y)}{P_R(x,y)}\right)\label{eq:f-div}\\
\nonumber&=\aver{f\!\left(\frac{P_F(x,y)}{P_R(x,y)}\right)}_F\;,
\end{align}
where $f:\mathbb{R}^+\to\mathbb{R}$ is a function to be specialized further in what follows. For the time being, it suffices to notice that for $f(r)=\ln r$ one recovers the Kullback-Leibler divergence~\cite{kullback1951}, i.e., the usual relative entropy, whereas for $f(r)=r^{\alpha}$ one recovers the family of Hellinger-R\'enyi divergences~\cite{liese-miescke}. We will come back to these particular choices at the end of this section. 

The forward-reverse ratio will be henceforth denoted
\ba\label{eq:frratio}
r(x,y)&=&\frac{P_F(x,y)}{P_R(x,y)}\,.
\ea
A priori, one should pay attention to the pairs $(x,y)$ such that this ratio is not well-defined. If necessary, this problem can be dealt with on a case-by-case basis, but there is no need to burden the notation already at this point. Besides, common sense demands (and our later definition will vindicate it) that if a pair $(x,y)$ is assigned a null probability under the forward process, it should be so also under the reverse process, and vice versa. Thus, ultimately, $r(x,y)$ will be ill-defined only for pairs $(x,y)$ that don't contribute to any average because $P_F(x,y)=P_R(x,y)=0$.

\subsection{$f$-fluctuation relations}

Eq.~\eqref{eq:f-div} suggests to interpret $f(r(x,y))$ itself as a \textit{random variable}, whose realization is denoted $\omega_F(x,y)$ for the forward process. The same random variable, when evaluated for the reverse process, is denoted $\omega_R(x,y)$. For consistency, these two must be the \textit{same function}, though evaluated on different arguments: more explicitly,
\begin{align}\label{eq:omegaFR}
\omega_F(x,y)&:=f(r(x,y))\nonumber\\
&\Downarrow\\
\omega_R(x,y)&:=f\!\left(\frac{1}{r(x,y)}\right)\;,\nonumber
\end{align}
simply due to the fact that when considering the reverse variable $\omega_R$, the roles of forward and reverse processes are exchanged and the ratio is thus inverted.

A Crooks-type fluctuation relation is a relation between the probability density functions $\mu$ of $\omega$ for the forward and the backward processes:
\ba
\mu_F(\omega)&=&\sum_{x,y}\delta(\omega_F(x,y)-\omega)P_F(x,y)\label{eq:muF}\\
\mu_R(\omega)&=&\sum_{x,y}\delta(\omega_R(x,y)-\omega)P_R(x,y)\nonumber\\
&=&\sum_{x,y}\delta(\omega_R(x,y)-\omega)\frac{1}{r(x,y)}P_F(x,y)\,.\label{eq:trivial}
\ea
If $f:\mathbb{R}^+\to \mathbb{R}$ is invertible, everything is well defined, and moreover there exists another function $g$ such that $f(1/r)=g(f(r))$, i.e., $\omega_R(x,y)=g(\omega_F(x,y))$. More explicitly, $g(u)=f(\frac{1}{f^{-1}(u)})$. Plugging these definitions into \eqref{eq:trivial}, we obtain
\begin{align*}
&\mu_R(u)\\
&=\sum_{x,y}\delta(\omega_R(x,y)-u)f^{-1}(\omega_R(x,y))P_F(x,y)\\
&=f^{-1}(u)\sum_{x,y}\delta(g(\omega_F(x,y))-u)P_F(x,y)\\
&=\frac{f^{-1}(u)}{|g'(g^{-1}(u))|}\sum_{x,y}\delta(\omega_F(x,y)-g^{-1}(u))P_F(x,y)
\end{align*}
and thus, with $u=g(\omega)$, we have a Crooks-type relation of the form
\ba\label{eq:Crooks0}
\mu_R(g(\omega))&=&\frac{f^{-1}(g(\omega))}{|g'(\omega)|}\,\mu_F(\omega)\,.
\ea Moreover, since $\int_{\mathbb{R}^+}\mu_R(g(\omega))|g'(\omega)|d\omega=\int_{\mathbb{R}^+}\mu_R(u)du=1$, there follows the corresponding Jarzynski-like relation \ba\label{eq:Jarlike0}
\aver{f^{-1}(g(\omega))}_F&=&1\,.
\ea

The above relation can be verified by unraveling our notations: since $\omega$ here is $\omega_F$, we have $g(\omega)=\omega_R$, and $f^{-1}(\omega_R)=1/r$. Thus $\aver{f^{-1}(g(\omega))}_F=\aver{1/r}_F=\sum_{x,y}\frac{1}{r(x,y)}P_F(x,y)=\sum_{x,y} P_{R}(x,y)=1$. Both this direct proof and the derivation from \eqref{eq:Crooks0} show that, from our perspective, the Jarzynski-like relation \eqref{eq:Jarlike0} expresses the normalization of the reverse process.

\subsection{Examples}

The usual choice made in the literature is \[\omega=f(r)=\frac{1}{z}\ln r\;,\] with $z\neq 0$. In this case, we have $f^{-1}(\omega)=e^{z\omega}$, $f(1/r)=-\frac{1}{z}\ln r$, and thus $g(\omega)=-\omega$. The relations \eqref{eq:Crooks0} and \eqref{eq:Jarlike0} become the familiar ones, namely,
\begin{align}\label{eq:FRgen}
\mu_R(-\omega)=e^{-z\omega}\mu_F(\omega)\;,
\end{align}
and
\begin{align*}
\aver{e^{-z\omega}}_F=1\,.
\end{align*}
Notice that one could have imposed the condition $\omega_R(x,y)=-\omega_F(x,y)$ from the start, in a way somehow reminiscent of an ``arrow-of-time'' variable. In our notation, this condition is equivalent to impose that the function $f$ satisfies $f(1/r)=-f(r)$, which leads to $f(r)=\frac{1}{z}\ln r$ for an arbitrary $z$.

As a second example, consider \[\omega=f(r)=r^{\alpha}\;,\] with $\alpha\neq 0$. Then $f^{-1}(\omega)=\omega^{1/\alpha}$, $f(1/r)=r^{-\alpha}$, and therefore $g(\omega)=1/\omega$. The fluctuation relations \eqref{eq:Crooks0} and \eqref{eq:Jarlike0} for random variables that satisfy $\omega_R=1/\omega_F$ are therefore
\ba\nonumber
\mu_R(1/\omega)=\omega^{2-1/\alpha}\mu_F(\omega)&\;\implies\;& \aver{\omega^{-1/\alpha}}_F=1\,.
\ea

Other examples may look exotic, but nonetheless possible: for instance, by choosing $\omega=f(r)=e^{\kappa r}$, with $\kappa\neq 0$, one obtains $\omega_R=e^{\kappa^2/\ln\omega_F}$, that is, $\ln\omega_R\ln\omega_F=\kappa^2$.



\section{Reverse process from Bayesian retrodiction}
\label{sec:Bayes}

In the previous section we assumed that forward and reverse processes were both given. However, when only the forward process is given, the reverse process should be \textit{derived} from it. In this section we explain how the reverse process can be derived from the forward process by applying the formalism of Bayesian retrodiction.

\subsection{Basics of Bayesian update}

We begin this section by reviewing the theory of Bayesian retrodiction. For simplicity, let us consider two random variables $X$ and $Y$ with states labeled by the indices $x$ and $y$, both taken from a finite set $\set{A}$. Let $P_{XY}(x,y)$ denote their joint distribution. If one's knowledge on $Y$ is updated to a definite value $y=y^*$, the Bayes--Laplace rule tells that the agents should update their belief on $X$ according to $P'_X(x)=P_{X|Y}(x|y^*)$. In other words, the joint probability is updated as
\ba
P'_{XY}(x,y)&=&P_{X|Y}(x|y)\,\delta_{y,y^*}\,.
\ea

The above formula, which constitutes the standard formulation of the Bayes--Laplace rule, is silent however about those situations, most common in real scenarios, in which the update does not result in a definite value $y^*$, but is itself described in terms of another probability distribution $P_Y'(y)$. In the face of such ``soft evidence'', the update should follow \textit{Jeffrey's conditioning}~\cite{jeffrey}
\ba\label{eq:jeffrey}
P'_{XY}(x,y)&=&P_{X|Y}(x|y)\,P'_Y(y)\,.
\ea
Jeffrey based this update on his ``rule of probability kinematics''. The coefficients $P_{X|Y}(x|y)$ are seen as defining a \textit{channel} that propagates the soft belief acquired about $y$ back onto $x$. It was later noticed that Jeffrey's update can also be obtained from Bayes--Laplace rule using Pearl's ``method of virtual evidence'' ~\cite{pearls,jaynes_2003,CHAN200567,jacobs-changing-mind}. From this viewpoint, the soft evidence on $Y$ is consequence of another \textit{definite} evidence on a ``virtual'' variable $Z$, that has no direct influence on $X$ (i.e.~$X\to Y\to Z$ forms a Markov chain). When $Z$ is updated to a definite value $z^*$, one updates the belief on $Y$ to $P'_{Y}(y)\equiv P_{Y|Z}(y|z)\,\delta_{z,z^*}$ and Eq.~\eqref{eq:jeffrey} is recovered.

\subsection{From update to retrodiction}

The Bayesian \textit{update} we just described becomes \textit{retrodiction} when the variables $X$ and $Y$ are given a diachronic meaning: $X$ represents the system's state at the initial time $t=0$, while $Y$ represents the system's state at the final time $\tau>0$.

As already noticed, if we know the \textit{forward} transition probability $P_{Y|X}(y|x)$, henceforth denoted as $\varphi(y|x)$, any prior knowledge $p(x)$ on $X$ can be forward-propagated using it. Hence, the \textit{forward process} will be
\ba
P_F(x,y)&=&p(x)\varphi(y|x)\label{eq:forward}
\ea
where $p(x)$ can be chosen arbitrarily. Now we want to define the \textit{reverse} transition probability $\hat{\varphi}(x|y)$, with the same functionality: it should be possible to use it to back-propagate onto $X$ any prior knowledge $q(y)$ that one obtains about $Y$. In other words, the \textit{reverse process} will be
\ba
P_R(x,y)&=&q(y)\hat\varphi(x|y)\label{eq:reverse}
\ea where $q(y)$ can be chosen arbitrarily. 

The reverse transition is obtained from Jeffrey's conditioning \eqref{eq:jeffrey}, but the knowledge of the forward transition alone is not enough \cite{watanabe65}. One needs to define a \textit{reference prior} $P_X(x)$. In the context of irreversibility, a natural choice is to set the reference prior equal to a steady (viz. invariant) state $\gamma(x)$, that is, a distribution such that
\[
\gamma(y)=\sum_x\gamma(x)\varphi(y|x)\;,
\]
for all $y\in\set{A}$. This choice coincides with what is customarily done in the theory of Markov chains when defining the reverse chain~\cite{norris_1997}.  Notice that, while every transition matrix possesses at least one steady state, the steady state may not be unique: in such a case, to any choice of a steady state there will correspond a different reverse transition, hence a different fluctuation relation.

Starting from a steady state, the \textit{reverse transition} $\hat\varphi(x|y)$ is defined by the relation $\gamma(y)\hat\varphi(x|y)=\gamma(x)\varphi(y|x)$. We see that $\gamma$ is an invariant distribution for the reverse process too (in other words: by choosing the steady state as prior, we define a reference process that does not distinguish between forward and reverse evolution). Implicitly restricting the analysis to the pairs $(x,y)$ with strictly positive weight (i.e., $\gamma(x)\varphi(y|x)>0$), we obtain
\begin{align}\label{eq:inversion}
\frac{\hat\varphi(x|y)}{\varphi(y|x)}=\frac{\gamma(x)}{\gamma(y)}\;.
\end{align} Plugging this together with Eqs.~\eqref{eq:forward} and~\eqref{eq:reverse} into \eqref{eq:frratio}, we find \ba
r(x,y)=\frac{p(x)\gamma(y)}{q(y)\gamma(x)}\;,\label{eq:ratioeq}
\ea that is, the forward-reverse ratio $r(x,y)$, which is the crucial quantity in the study of irreversibility as noticed in Section~\ref{sec:FT}, depends on the stochastic transition $\varphi(y|x)$ only through its steady state $\gamma$.


\section{Quantum inside: recovering quantum retrodiction}
\label{sec:qinside}

When the channel is realised by a quantum process, we proceed to show that the formalism of Jeffrey's conditioning is automatically compatible with the formalism of quantum retrodiction~\cite{barnett-pegg-jeffers,fuchs2002quantum,Leifer-Spekkens,bergou-symm-retro}.

Here quantum mechanics enters the picture assuming that the stochastic transition $\varphi(y|x)$ involves an inner quantum ``mechanism'': to each input $x\in\mathcal{A}$ is associated an input state (density matrix) $\rho_0^x$, that is later propagated to $\rho_\tau^x=\mE(\rho_0^x)$ via a completely positive trace-preserving (CPTP) linear map $\mE$, and finally measured using a positive operator-valued measure (POVM) with outcomes $y\in\mathcal{A}$ and elements $\Pi_\tau^y$. The subscripts $0$ and $\tau$ are used to denote, respectively, an initial time $t=0$ and a final time $t=\tau>0$. With these notations,
\ba
\varphi(y|x)&=&\textrm{Tr}[\Pi_\tau^y\ \mE(\rho_0^x)]\,.\label{eq:channel}
\ea
In the above equation, the density matrix $\rho_0^x$ is meant to encode all the relevant degrees of freedom needed to represent the $x$-th experimental setup. Hence, it can account for processes in which, for example, there is no clear distinction between system and environment due to the presence of initial correlations~\cite{pechukas,alicki-comment,stelmachovic-buzek,sudarshan,buscemi-NCPTP,lidar-dominy}. In all such cases, $\rho_0^x$ will include not only the degrees of freedom typically associated with the system, but also those associated with the environment.

The expression of $\hat{\varphi}$ in the quantum formalism is immediately obtained from \eqref{eq:channel}. By introducing the state $\gamma_0=\sum_x\gamma(x)\rho^x_0$, that we assume invertible (otherwise we can restrict the analysis to the subspace where $\gamma_0>0$), one gets the much more evocative expression
\begin{align}
\hat\varphi(x|y)&=\Tr[\Theta^x_0\ \hat\mE(\sigma^y_\tau)]\label{eq:q-retro}
\end{align}
which is the Born rule for the POVM elements
\begin{align}\label{eq:retro-POVM}
\Theta^x_0:=\gamma(x)\frac{1}{\sqrt{\gamma_0}}\rho^x_0\frac{1}{\sqrt{\gamma_0}}\;,
\end{align}
the normalized states
\begin{align}\label{eq:retro-states}
\sigma^y_\tau:=\frac{1}{\gamma(y)}\sqrt{\mE(\gamma_0)}\ \Pi^y_\tau \ \sqrt{\mE(\gamma_0)}\;,
\end{align}
and the reverse quantum channel~\cite{petz,barnum-knill,crooks-reversal}
\begin{align}\label{eq:retro-channel}
\hat\mE(\cdot):=\sqrt{\gamma_0}\ \mE^\dagger\left[\frac{1}{\sqrt{\mE(\gamma_0)}}(\cdot)\frac{1}{\sqrt{\mE(\gamma_0)}}\right]\ \sqrt{\gamma_0}\;,
\end{align}
$\mE^\dag$ being the trace-dual of $\mE$, defined by the relation $\Tr[\mE^\dag(X)\ Y]=\Tr[X\ \mE(Y)]$ for all operators $X$ and $Y$. We assume $\mE(\gamma_0)>0$, so that $\hat\mE$ is a CPTP linear map defined everywhere, and thus physically realizable\footnote{Even if $\mE(\gamma_0)$ is not invertible, $\hat\mE$ can always be extended to a CPTP map defined everywhere and, thus, physically realizable~\cite{wilde_2013}.}.

Just as Eq.~\eqref{eq:inversion} is the \textit{classical retrodiction} for $\varphi(y|x)$, its quantum description given in Eq.~(\ref{eq:q-retro}) constitutes the \textit{quantum retrodiction} of Eq.~\eqref{eq:channel}, in perfect agreement with previous literature on quantum retrodiction~\cite{barnett-pegg-jeffers,fuchs2002quantum,Leifer-Spekkens,bergou-symm-retro}.


\section{Classical Thermodynamics of Retrodiction}
\label{sec:examples}


In this Section, we show how several important classical fluctuation relations can be recovered using the retrodictive approach.


\subsection{Doubly-stochastic transitions, and classical Hamiltonian dynamics}\label{subsec:doubly-stoch}

We start by noticing that the condition
\begin{align}\label{eq:retro-ds}
\hat\varphi(x|y)=\varphi(y|x)
\end{align}
holds if and only if the transition is doubly-stochastic, viz.~if in addition to the compulsory normalisation $\sum_y\varphi(y|x)=1$, it also holds $\sum_x\varphi(y|x)=1$. Indeed, comparing with~\eqref{eq:inversion}, we see that \eqref{eq:retro-ds} holds if and only if the steady state can be chosen as the uniform distribution $\gamma(x)\propto 1$. It is then trivial to check that doubly stochastic channels always admit such a steady state, while channels that are not doubly stochastic do not have a uniform steady state.

A particularly important case of doubly stochastic transition is \textit{classical Hamiltonian dynamics}, which is deterministic, viz.~there is a one-to-one correspondence between the initial state $x$ and the final state $y\equiv x'$. The evolution $x\to x'$ is then represented by the transition $\varphi(x'|x)$, with $\varphi(x'|x)=1$ if $x'$ is the final state corresponding to the initial state $x$, and $\varphi(x'|x)=0$ otherwise. Thus, for any classical Hamiltonian process one can choose $\gamma$ as uniform.

For instance, the scenario considered by Bochkov and Kuzovlev \cite{bochkov-kuzovlev-1977} is of this type. They specify a class of driven Hamiltonians such that, on any given forward trajectory $x\rightarrow x'$, the non-driven term $H_0$ satisfies $H_0(x')=H_0(x)+E(x,x')$, with $E$ determined by the driving protocol. By choosing thermal priors $p(x)\propto e^{-\beta H_0(x)}$ and $q(x')\propto e^{-\beta H_0(x')}$, with $\beta=1/k_BT$, they obtain
\ba
r(x,x')&=&\frac{\varphi(x'|x)p(x)}{\hat\varphi(x|x')q(x')}=\frac{p(x)}{q(x')}\,=\,e^{-\beta E(x,x')}\,.
\ea Since the process is deterministic, the statistics (and in particular the fluctuation relations) carry over from the initial conditions to the whole trajectory. From this observation, rich physical consequences follow: one can derive many-point relations, relations for the currents (e.g.~Onsager), linear response results (in the limit when the driving forces go to zero), etc.~\cite{bochkov-kuzovlev-1977}. As we presented it, all this can be seen as starting with retrodiction.

Another example that fits in this subsection is that of a generic time-dependent Hamiltonian, but with \textit{microcanonical} priors instead of the most frequently used thermal ones~\cite{morillo-2008-microcanonical}. Staying with a discrete alphabet again (the generalisation follows immediately), such priors read: $p(x)=N(E)^{-1}$ if $E_x= E$ (and zero otherwise) and $q(x')=N(E')^{-1}$ if $E_{x'}= E'$ (and zero otherwise), where $N(E)$ and $N(E')$ are the degeneracies of the two levels. Thus, for the processes with non-zero probability, the forward-reverse ratio reads
\begin{align}
r(x,x')&=\frac{p(x)}{q(x')}=\frac{N(E')}{N(E)}\nonumber\\
&=e^{(S(E')-S(E))/k_B}\;,\label{eq:micro}    
\end{align}
where $S(E):=k_B\ln N(E)$ coincides with Boltzmann's entropy formula.

\subsection{Classical Hamiltonian system-reservoir interactions}

We consider now a system composed of two subsystems. For notational purposes, let us denote the microstates of the first system at initial and final time by the labels $x$ and $x'$, respectively; and those of the second system analogously by the labels $w$ and $w'$. For the purpose of the present example, all labels belong to discrete sets. If the joint evolution is Hamiltonian, we can borrow from the previous discussion and conclude that \begin{align}\label{eq:retro-jarz2000}
\hat\varphi(x,w|x',w')=\varphi(x',w'|x,w)\;.
\end{align}
To construct the forward and reverse processes \eqref{eq:forward} and~\eqref{eq:reverse}, one should specify the prior distributions $p(x,w)$ and $q(x',w')$.

This notation opens the possibility of \textit{coarse-graining} over one of the subsystems. For definiteness, we describe the particular case studied in~\cite{jarz2000}. In this narrative, the second system consists of one or several \textit{heat reservoirs}. It is rather natural therefore to stipulate that:
\begin{enumerate}
	\item Both priors are in product form, i.e., $p(x,w)=p(x)P(w)$ and $q(x',w')=q(x')Q(w')$;
	\item The reservoirs' priors $P(w)$ and $Q(w')$ are thermal distributions at inverse temperature $\beta^{-1}=k_BT$, i.e., $P(w)\propto e^{-\beta E_w}$ and $Q(w')\propto e^{-\beta E_{w'}}$, where $E_w$ and $E_{w'}$ are the energies of the reservoir's microstates $w$ and $w'$. This condition implies
\begin{align}\label{eq:jarz2000-balance}
\frac{P(w)}{Q(w')}=e^{\beta(E_{w'}-E_w)}=:e^{\Delta S/k_B}
\end{align} where $\Delta S$ is the entropy generated.
\end{enumerate}

With these two assumptions on the prior distributions, one can compute the following marginal conditional probability:
\begin{align}\label{eq:jarz2000-forward}
\varphi(x',\Delta S|x)=\sum_{w,w':E_{w'}-E_w=T\Delta S}\varphi(x',w'|x,w)P(w)\;.
\end{align}
The above represents the forward transition probability that the system, if starting in microstate $x$, will end up in microstate $x'$ generating in the process an amount of entropy equal to $\Delta S$. Notice that
the coarse-grained transition $\varphi(x',\Delta S|x)$ computed in~\eqref{eq:jarz2000-forward} only depends on the reservoir's prior $P(x)$. In other words, while the reservoir's prior distribution $P(x)$ is fixed, the system's prior remains arbitrary.

Analogously, but starting from the retrodicted transition~\eqref{eq:retro-jarz2000}, we obtain
\begin{align}
&\hat\varphi(x,-\Delta S|x')\nonumber\\
&=\sum_{w,w':E_{w}-E_{w'}=-T\Delta S}\hat\varphi(x,w|x',w')Q(w')\nonumber\\
&=\sum_{w,w':E_{w}-E_{w'}=-T\Delta S}\varphi(x',w'|x,w)Q(w')\label{eq:passaggio0}\\
&=\sum_{w,w':E_{w'}-E_w=T\Delta S}\varphi(x',w'|x,w)P(w)\frac{Q(w')}{P(w)}\nonumber\\
&=\sum_{w,w':E_{w'}-E_w=T\Delta S}\varphi(x',w'|x,w)P(w)e^{-\Delta S/k_B}\label{eq:passaggio}\\
&=e^{-\Delta S/k_B}\varphi(x',\Delta S|x)\nonumber\;,
\end{align}
where in~\eqref{eq:passaggio0} we used relation~\eqref{eq:retro-jarz2000}, while in~\eqref{eq:passaggio} we used relation~\eqref{eq:jarz2000-balance}. The above represents the reverse transition probability that the system, if starting in microstate $x'$, will end up in microstate $x$ generating in the process an amount of entropy equal to $-\Delta S$. Summarizing, we have recovered
\ba
\frac{\varphi(x',\Delta S|x)}{\hat\varphi(x,-\Delta S|x')}=e^{\Delta S/k_B}\;,\label{eq:jarz2000}
\ea which is the main result of~\cite{jarz2000}. It is worth stressing that, as written, $\Delta S$ plays the role of an additional random variable coarse-graining the reservoir's microstates: that is, $\varphi(x',\Delta S|x)$ describes a transition from input $x$ to output $(x',\Delta S)$. Thus, the object at the denominator in~\eqref{eq:jarz2000} is not the complete Bayesian reverse of the numerator, but rather a \textit{partial} (viz. ``hybrid''~\cite{jarz2000}) reversal.

\subsection{Retrodiction in stochastic thermodynamics}

However desirable it would be to derive everything from the Hamiltonian dynamics of a closed (possibly composite) system, information about the latter is often lacking. This is when the approach based on Bayesian retrodiction really shows its power: it allows one to make inferences based on whatever partial information is available.

As a first example of this type, we consider the stochastic thermodynamics setting that led to Crooks' theorem~\cite{crooks-theorem}. A general process in discrete-time stochastic thermodynamics is modeled as a sequence of external driving protocols (the \textit{work steps}) alternating with periods during which the system is allowed to equilibrate with an ideal heat bath (the \textit{relaxation steps})~\cite{crooks-theorem,thermo-prediction}. The changes in the system's internal energy during each work step are counted as work done on the system, while the changes happening during each relaxation steps are counted as heat absorbed by the system. For simplicity, we consider here just a two-step process: one work step followed by one relaxation step.

Let us denote the system's initial state by $x$, and let $E_x$ be the system's initial energy. The system is (deterministically) driven to another energy $E'_x$. This constitutes the work step. The relaxation step, which is not deterministic, is modeled using a transition conditional probability $\varphi(y|x)$, where $y$ labels the system's state after the relaxation step. Assuming that during the relaxation step the system's Hamiltonian does not change, the system's final energy, after the relaxation step is over, is $E'_y$.

By definition, an invariant distribution for $\varphi(y|x)$ is the thermal distribution $\gamma(x)\propto e^{-\beta E'_x}$. This, via~\eqref{eq:inversion}, leads to the retrodicted transition
\begin{align}
\hat\varphi(x|y)&=\frac{\gamma(x)}{\gamma(y)}\varphi(y|x)\nonumber\\
&=e^{\beta(E'_y-E'_x)}\varphi(y|x)\nonumber\;.
\end{align}
Choosing as priors the thermal distributions, that is, $p(x)= e^{\beta(F- E_x)}$ and $q(y)= e^{\beta(F'- E'_y)}$, with $F,F'$ the corresponding Helmholtz free energies, the ratio~\eqref{eq:ratioeq} becomes
\begin{align*}
r(x,y)&=\frac{\varphi(y|x)p(x)}{\hat\varphi(x|y)q(y)}\\
&=e^{-\beta (E'_y-E'_x)}e^{\beta(E'_y-E_x-\Delta F)}\\
&=e^{\beta(E'_x-E_x-\Delta F)}=e^{\beta (W-\Delta F)}\;,
\end{align*}
where in the last passage we used the assumption that the system's internal energy change during the work step is work $W$ done on the system. The fluctuation relation~\eqref{eq:FRgen} then immediately gives
\[
\frac{\mu_F(W)}{\mu_R(-W)}=e^{\beta (W-\Delta F)}\;,
\]
in accordance with~\cite{crooks-theorem}. Notice that in our retrodictive derivation we did not require any particular condition (such as the condition of detailed balance) for the relaxation step $\varphi(y|x)$, apart from it preserving the thermal distribution, which is implicit in the definition of ``relaxation''.


\section{Quantum Thermodynamics of Retrodiction}
\label{sec:exquantum}

Our definition of the reverse quantum channel based on Bayesian inversion, presented in Section~\ref{sec:qinside}, accommodates any state preparation $\{\rho_0^x:x\in\mathcal{A}\}$, any quantum channel $\mathcal{E}$, and any final measurement $\{\Pi^y_\tau:y\in\mathcal{A}\}$. It hence contains, as a very special case, the conventional setup of Tasaki's two-measurement thermodynamics of closed driven quantum systems~\cite{tasaki2000jarzynski}. But it also contains, and resolves, some quantum setups that have been considered problematic in the past due to the supposed lack of a well-defined reverse process (see e.g.~\cite{Albash2013,rastegin-zycz-2014,Funo2018}). Where physical intuition fails to envisage a ``natural'' inversion of the physical process, consistent logical inference (viz. Bayesian inversion) comes to the rescue.

\subsection{Two-measurement setup for closed driven quantum systems}

The paradigm for quantum fluctuation relations is provided by Tasaki's \textit{two-measurement setup}~\cite{tasaki2000jarzynski}. Here, a $d$-level quantum system is prepared in the state $\rho_0$ but immediately subjected to a von Neumann projective measurement of the initial Hamiltonian $H_0=\sum_x\epsilon_x|\epsilon_x\>\<\epsilon_x|$. The system, now collapsed onto $|\epsilon_x\>$ after observation of $\epsilon_x$, next undergoes a perfectly adiabatic work protocol: its Hamiltonian is driven from $H_0$ to $H_\tau=\sum_y\eta_y|\eta_y\>\<\eta_y|$, but the system is otherwise perfectly isolated from the surrounding environment. At the end of the driving protocol, during which only mechanical work has been exchanged with the system, the system is subjected to a second energy measurement, this time of the final Hamiltonian $H_\tau$.

Let us analyse Tasaki's setup with our tools. Denoting the unitary evolution resulting from the driving protocol by $U_{0\to \tau}$, the forward (predictive) transition probability is given by
\begin{align}\label{eq:unitary-forward}
\varphi(y|x)=\Tr[U_{0\to \tau} |\epsilon_x\>\<\epsilon_x| U_{0\to \tau}^\dagger\ |\eta_y\>\<\eta_y|]\;.
\end{align}
It is easy to verify that $\varphi(y|x)$ is doubly stochastic (see Subsection~\ref{subsec:doubly-stoch}). This is a consequence of the following three facts: (i) that the underlying quantum process is unitary; (ii), that $\Tr[|\epsilon_x\>\<\epsilon_x|]=\Tr[|\eta_y\>\<\eta_y|]=1$; and (iii), that $\sum_x|\epsilon_x\>\<\epsilon_x|=\sum_y|\eta_y\>\<\eta_y|=\openone$. As noted in Subsection~\ref{subsec:doubly-stoch}, the reverse (retrodictive) transition is given by
\begin{align}\label{eq:bistochastic-reverse}
\varphi(y|x)=\hat\varphi(x|y)\;.
\end{align}
At the same time, the corresponding quantum retrodiction, given in Eqs.~\eqref{eq:retro-POVM}-\eqref{eq:retro-channel}, is in perfect agreement with a narrative involving time-reversals, as in e.g.~\cite{sagawa2012second}: one first prepares the eigenstates of $H_{\tau}$, then evolves them ``backwards in time'' using $U_{0\to \tau}^\dag$, and finally measures $H_0$. While such a narrative seems simple and appealing to intuition, it is not directly \textit{operational}, as the actual implementation of time-reversals is far from straightforward~\cite{quintino-murao-reversing-unitaries,gilyn2020quantum}. Moreover, as we will see in what follows, when the evolution between the two energy measurements is not Hamiltonian, a naive argument using time-reversal can lead to inconsistencies.

In any case, with Eq.~\eqref{eq:bistochastic-reverse} at hand, fluctuation relations can be easily derived. One only needs to specify the two priors $p(x)$ and $q(y)$, which in the two-measurement setup is tantamount to specifying two quantum states $\rho_0$ and $\sigma_\tau$, since from them we have $p(x)=\<\epsilon_x|\rho_0|\epsilon_x\>$ and $q(y)=\<\eta_y|\sigma_\tau|\eta_y\>$. Simply by choosing $\rho_0$ as the thermal state for $H_0$ and $\sigma_\tau$ as the thermal state for $H_\tau$ (by assuming that the temperature is the same), we find that
\begin{align*}
r(x,y)&=\frac{\varphi(y|x)p(x)}{\hat\varphi(x|y)q(y)}\\
&=\frac{p(x)}{q(y)}\\
&=e^{\beta (F_0-\epsilon_x)}e^{-\beta(F_\tau-\eta_y)}\\
&=e^{\beta(\eta_y-\epsilon_x-\Delta F)}\\
&=e^{\beta (W-\Delta F)}\;,
\end{align*}
where in the final passage we identified the system's energy difference as work $W$ done on the system, due to the assumption of adiabaticity during the driving protocol. From here, fluctuation relations analogous to Crooks' theorem and Jarzynski's equality quickly follow.

\subsection{Two-measurement setup for general quantum channels, and nonequilibrium potentials}

Tasaki's setup invites various generalizations. Several references~\cite{Albash2013,Rastegin_2013,rastegin-zycz-2014,GooldModiPater2015} have considered the situation in which the initial and final measurements are as in Tasaki's arrangement but the unitary evolution is replaced by a general CPTP linear map $\mE$, leading to the forward transition
\begin{align}
    \varphi(y|x)=\Tr[\mE(|\epsilon_x\>\<\epsilon_x|)\ |\eta_y\>\<\eta_y|]\,.\label{eq:epseta}
\end{align}
If the linear map $\mE$ is not unital, that is, if it does not preserve the unit matrix (viz. $\mE(\openone)\neq\openone$), then the above transition probability is not doubly stochastic in general. For the retrodictive transition, this means that $\hat\varphi(x|y)\neq\varphi(y|x)$, and indeed Eq.~\eqref{eq:inversion} contains the extra factor $\gamma(x)/\gamma(y)$ coming from Bayes--Laplace rule.

Nonetheless, a common approach prescribes to stay put with Eq.~\eqref{eq:bistochastic-reverse} and work with the \textit{non-normalized} conditional distribution $\tilde\varphi(x|y):=\varphi(y|x)$ instead. Obviously, $\tilde\varphi(x|y)$ does not admit a realization as in \eqref{eq:q-retro}, simply because it is not a well-formed conditional distribution. Nonetheless, it is still possible to construct a ``ratio''
\[
\tilde r(x,y)=\frac{\varphi(y|x)p(x)}{\tilde\varphi(x|y)q(y)}\equiv\frac{p(x)}{q(y)}
\]
and formally go through all the calculations as in the normalized case. However, as a result of considering a reverse ``process'' that is not properly normalized, the resulting Jarzynski--like relation does not average to 1 as one would like (and as it happens in~\eqref{eq:Jarlike0}), but to $\sum_{x,y}\tilde\varphi(x|y)q(y)$. This quantity is known as the \textit{efficacy}~\cite{Albash2013}, but we recognize it here as a mathematical artifact arising from an ill-defined reverse transition.

The formalism presented in this work allows a simple treatment of the two-measurement process also for arbitrary quantum channels. The forward transition~\eqref{eq:epseta} coincides with Eq.~\eqref{eq:channel} for $\rho_0^x:=|\epsilon_x\>\<\epsilon_x|$ and $\Pi^y_\tau:=|\eta_y\>\<\eta_y|$. Let $\gamma(x)$ be an invariant distribution for $\varphi(y|x)$. Further, let predictive and retrodictive prior distributions be thermal distributions for some energy levels $\epsilon_x$ and $\eta_y$, so that $p(x)\propto e^{-\beta\epsilon_x}$ and $q(y)\propto e^{-\beta\eta_y}$. The corresponding initial and final Hamiltonians are defined as $\sum_x\epsilon_x|\epsilon_x\>\<\epsilon_x|$ and $\sum_y\eta_y|\eta_y\>\<\eta_y|$, respectively.

The ratio is defined as usual, that is,
\begin{align*}
&r(x,y)=\frac{\varphi(y|x)p(x)}{\hat\varphi(x|y)q(y)}=\frac{p(x)}{q(y)}\frac{\gamma(y)}{\gamma(x)}\\
&=e^{\beta (F_0-\epsilon_x-\ln\gamma(x))}e^{-\beta(F_\tau-\eta_y-\ln\gamma(y))}\\
&=e^{\beta(\Delta E-\Delta\Phi-\Delta F)}\;,
\end{align*}
where now we find an extra term $\Delta\Phi:=\Phi_y-\Phi_x:=\frac{1}{\beta}\ln\gamma(x)-\frac{1}{\beta}\ln\gamma(y)$. This term is understood as the difference of a \textit{nonequilibrium potential} that adds to the difference of equilibrium free energy $\Delta F:=F_\tau-F_0$. Finally, introducing a total stochastic entropy production $\Sigma:=
\beta(\Delta E-\Delta\Phi-\Delta F)$, one easily obtains the detailed relation
\begin{align*}
\frac{\mu_F(\Sigma)}{\mu_R(-\Sigma)}=e^{\Sigma}\;,
\end{align*}
and the corresponding integral relation $\aver{e^{-\Sigma}}_F=1$. In this way, we can see how a correct application of the Bayes--Laplace inversion formula automatically takes into account nonequilibrium potentials. These, usually introduced as corrections~\cite{hatano-sasa-2001, manzano-fluctuations-unit-efficacy}, are now recognizable as the remnants of Bayesian inversion.

\section{Conclusion}

Studies of irreversibility rely on the comparison between a forward physical process and its reverse. We have proposed to define the latter as a form of Bayesian retrodiction~\eqref{eq:inversion}. We showed that, on the one hand, this definition matches the one used in the derivation of canonical results like Jarzynski's equality, Crooks' theorem, and Tasaki's two-measurement fluctuation theorem for closed driven quantum systems. On the other hand, it also applies to situations in which a reverse process was supposed to be lacking. As a by-product, various modifications, like non-unit efficacies or nonequilibrium potentials, are given a simple explanation as shadows of Bayes--Laplace inversion. 

Logical inference thus emerges as a powerful tool to supplement or replace physical intuition, whenever this seems hard to obtain. Clearly, the present approach opens up the possibility for various developments in statistical mechanics and beyond. In particular, an important development is to prove quantum retrodiction as the logical foundation of a fully quantum fluctuation theorem, such as those posited in~\cite{alhambra-fluct-work,aberg-quantum-fluct,manzano-PRX,kwon-kim}, but we leave this for future research.

\section*{Acknowledgments}

 The authors thank Paul Riechers for insightful comments. F.B. acknowledges support from the Japan Society for the Promotion of Science (JSPS) KAKENHI, Grants Nos.19H04066 and 20K03746, and from MEXT Quantum Leap Flagship Program (MEXT Q-LEAP), Grant Number JPMXS0120319794. V.S. acknowledges support from the National Research Foundation and the Ministry of Education, Singapore, under the Research Centres of Excellence programme.

\bibliographystyle{alphaurl}
\bibliography{references}
\end{document}